\documentclass[preprintnumbers,amsmath,amssymb,onecolumn]{revtex4}
\usepackage[driverfallback=dvipdfm]{hyperref}
\usepackage{color}
\usepackage{graphicx}
\usepackage{makeidx}
\usepackage{bm}
\usepackage[title]{appendix} 
\usepackage{tabu}
\usepackage{float}
\usepackage{chemarr}

\begin{document}
\title{\large Epidemiological parameter sensitivity in Covid-19 dynamics and estimation}
\author{Jyoti Bhadana, R.K. Brojen Singh}
\email{brojen@jnu.ac.in}
\affiliation{School of Computational $\&$ Integrative Sciences, Jawaharlal Nehru University, New Delhi-110067, India}

\begin{abstract}
{\noindent}Covid-19 is one of the most dreaded pandemics/epidemics in the world threatening the human population. The dynamics of this pandemic is quite complicated and prediction of pandemic states often fails. In this work, we study and correlate the SIR epidemiological model with the ongoing pandemic and found that pandemic dynamics and states are quite sensitively dependent on model parameters. The analysis of the exact parametric solution of the deterministic SIR model shows that the fixed points ($g^*$) depend on the SIR parameters, where, if $g^*\langle\frac{\gamma}{\beta x_0}$ then $g^*$ is stable and the pandemic can be controlled, whereas, if $g^*\rangle\frac{\gamma}{\beta x_0}$ then $g^*$ is unstable indicating active pandemic state, and $g^*=0$ corresponds to an endemic state. The dynamics show asymptotic stability bifurcation. The analytical solution of the stochastic SIR model allows to predict endemic time $t_E\propto\frac{1}{\gamma}$ and mean infected population becomes constant for small values of $\gamma$. The noise in the system even can modulate pandemic dynamics. Hence, we estimated the values of these parameters during lockdown periods using the ABC SSA algorithm and found strong dependence on control strategy, namely, lockdown and follows power-law behaviors. The numerical results show strong agreement with the real data and have the positive effect of social distancing or lockdown in the pandemic control.\\

\end{abstract}

%PACS number(s): 71.23, 72.15.R, 73.50
%%%%%%%%%%%%%%%%%%%%%%%%%%%%%%%%%%%%%%%%%%%%%%%%%%%%%%%%%%%%%%%%%%%%%%%%%%%%%%%%%%%%%%%%%%%%%%%%%%%%%%
\maketitle

\vskip 0.7cm
{\noindent}\textbf{\large Introduction}\\
\\
{\noindent}Covid-19 pandemic is complex multifaceted dynamics leading to difficulty in disease state prediction and control. The first case of the ongoing Covid-19 pandemic was identified in Wuhan, China, in December 2019, and spread over across the world \cite{Holshue}. Based on the peculiar nature of fast-spreading pneumonia which causes mass deaths across the world and global threat to the human population, the World Health Organization (WHO) declared it as a pandemic in March 2020 for special attention to prevent this disease \cite{WHO}. The first Covid-19 positive case in India was identified on 30 January 2020, and now India is reported to be the second-highest number of confirmed Covid-19 positive cases all over the world \cite{India}. As a control strategy for this disease, the Indian Government applied the lockdown scheme to prevent the pandemic under India's prime minister’s order. The first lockdown was placed for 21 days from 25 March 2020 to 14 April 2020, followed by some rules and regulations. The growth rate of the pandemic reduced, but after 6 April, positive cases increased with double rate \cite{Sandhya}. So the government decided to extend the lockdown. The lockdown extended for 19 days till 3 May 2020 on 14 April with some relaxation. Later the lockdown extended for two weeks until 17 May by dividing all the districts into three zones based on the spread dynamics - green, red, and orange. On 17 May, the lockdown was further extended till 31 May. On 30 May, the Unlock scheme came with relaxations applied accordingly. Even though a mass vaccination campaign was announced and continues to prevent this pandemic \cite{Bagcchi}, the second wave attacked \cite{Ranjan}. However, still, this pandemic is not fully controlled and why it is difficult to control this pandemic is still an open question.\\

{\noindent}There have been quite many mathematical models to study the spreading dynamics of the Covid-19 and control strategies \cite{Rahimi}. A networked dynamic metapopulation model was presented to show the critical epidemiological characteristics of Covid-19 using the SIR model \cite{Li}. On the other hand, based on mobile data counts, the risk model was built to describe the spread of Covid-19 statistically and tracking \cite{Jia}. Again, various other stochastic extensions of the SIR models have also been proposed to understand disease dynamics for prevention \cite{Brauer}. Even though these studies provide important characteristics and dynamics of the Covid-19 pandemic, still this pandemic could not able to control properly.\\

{\noindent}In this work, we propose the need for analysis of the sensitivity of pandemic parameters on disease spreading dynamics which could be one potential strategy to correlate with the disease control mechanism. For this one needs to estimate the values of model parameters on various pandemic states triggered by adopted government policies to control the pandemic. The estimated parameter values can be used for the prediction of disease states. The idea of parameter estimation is based on Bayesian inference in which the prior belief and then concluding the distribution of parameters rather than predicting the single point value is the desirable property \cite{Box}. Exact Bayesian inference is tractable only for a small set of parametric distributions \cite{Worden}. Parameter estimation is a computational approach for the parameters under available observations. Some unpredictable fluctuations of the observations may differ from one experiment to another under the same existing conditions called nonsystematic errors—the effective way of describing these kinds of fluctuations by model observations stochastically. The function of observations used to compute the parameters is called the estimator. If the fluctuations of the observations were absent, the estimator would produce exact values for the parameters \cite{Van}. Bayesian inference methods described in epidemiological modeling are generally based on Markov Chain Monte Carlo (MCMC) process \cite{Hamra}. These Bayesian inference methods depend on the likelihood functions. The compatibility for a given model depends on the probability of observed data for a given set of parameters. Due to this incompatibility, epidemiology models are restricted. So new methods were developed that do not require the likelihood functions for parameter inference. Such algorithms for “likelihood-free inference” have been developed. Such approaches have been successfully applied to a variety of domains including evolution and ecology \cite{Beaumont,Mondal}, cosmology \cite{Weyant}, econometrics \cite{Calvet}, cognitive science \cite{Kang}, systems biology \cite{Liepe}, and biochemistry \cite{Tomczak}. The parameters are estimated by using the ABC method that is based on sequential Monte Carlo (SMC) for dynamical models \cite{Toni}. \\

{\noindent}The study of epidemiological models to understand Covid-19 dynamics deterministically or stochastically generally use a fixed value of the parameters involved in the respective models \cite{Adak}. Due to some of the complex peculiar properties of this pandemic such as the Covid-19 virus spreads quite fast \cite{Wang}, rate transmission is high causing an abrupt change in positive cases \cite{Novel} etc, in a country like India with a vast population, it is hard to prevent its spread. Indian Government's lockdown policy improves in recovery rates as compared to other countries \cite{Siettos} still the pandemic is not under control. In this process of prediction and understanding pandemic states, the reproduction number $R_0$ has been used as an indicator \cite{Siettos}.
In this paper, we used the classic SIR (susceptible-infected-recovery) model to study parameter sensitivity on pandemic/epidemic dynamics with special reference to the Covid-19 pandemic. Due to their sensitiveness to complicated dynamics, the estimated parameters can be used not only in describing complicated disease dynamics and states but also in forecasting pandemic fate. We picked up lockdown as a control strategy that triggers changes in the topology of the dynamics via changes in the parameter values. We propose that this sensitiveness of the parameter values in the pandemic needed to be considered and systematically implemented in the Covid-19 models for the proper and accurate pandemic forecast.  
%%%%%%%%%%%%%%%%%%%%%%%%%%%%%%%%%%%%%%%%%%%%%%%%%%%%%%%%%%%%%%%%%%%%%%%%%%%%%%%%%%%%%%%%%%%%%%%%%%%%%%%%%%%%%%%%%

\vskip 0.7cm
{\noindent}\textbf{\large Sensitivity of the pandemic dynamics on disease parameters}\\\\
{\noindent}The epidemiological model which can study disease dynamics in any epidemic or pandemic is the classic SIR (susceptible-infectious-recovered) model \cite{Kermack} with $x\rightarrow S/N,y\rightarrow I/N,z\rightarrow R/N$ ($N$ is the population in the demographic regime) as described by,
\begin{eqnarray}
\label{sir}
\frac{d\bf{\Lambda}}{dt}=\bf{\mathcal{F}(\Lambda)};~\bf{\Lambda}=\left[\begin{matrix}x\\y\\z \end{matrix}\right];~\bf{\mathcal{F}}=\left[\begin{matrix}-\beta xy\\\beta xy-\gamma y\\\gamma z \end{matrix}\right];~~~\forall x,y,z\in\mathbb{R}
\end{eqnarray} 
where, $\beta$ is the infection rate and $\gamma$ is the recovered rate which can related to the reproduction number $R_0$ by $\gamma=\frac{\beta}{R_0}$ \cite{Hethcote}. The analytical solution of this equation \eqref{sir} can be obtained by the following the procedure described by Harko et. al. \cite{Harko}. In this procedure, the coupled ordinary differential equations (ODE) in equation \eqref{sir} are converted to an equivalent single second order  ordinary differential equation in $z$, and then with the variable transformation $g=e^{-\frac{\beta}{\gamma}z}$, the ODE in $z$ is transformed to a simplified second order ODE in $g$. Then by defining a function $f=\frac{dt}{dg}$, this ODE in $g$ can be reduced to a type of Bernoulli differential equation which can be solved, and the $z-$dependent solution for initial condition $[x_0,y_0,z_0]^T$ are,
\begin{eqnarray}
x=x_0g(z);~y=\frac{\gamma}{\beta}ln[g(z)]-x_0g(z)+x_0+y_0;~z=-\frac{\gamma}{\beta}ln[g(z)]+z_0
\end{eqnarray}
where, the implicit function $g(z)$ can be calculated from the following ODE \cite{Harko,Toda},
\begin{eqnarray}
\label{geq}
\frac{dg}{dt}=g[\beta x_0(g-1)-\gamma ln(g)-\beta y_0]=F(g)
\end{eqnarray}
The analysis of this equation \eqref{geq} was done systematically in order to connect with Covid-19 pandemic behavior \cite{Toda}. Our study is the extension of the works of Harko et. al. \cite{Harko} and Toda \cite{Toda} to understand the sensitivity of the model parameters on pandemic behavior.\\

\noindent\textbf{Lemma 1:} \textit{The pandemic dynamics described by equation \eqref{geq} has two types of fixed points:
\begin{itemize}
\item The fixed point $g^*=0$ provides a stable fixed point which corresponds to "endemic" condition where, $z^*\rightarrow\infty$ as $t\rightarrow\infty$.
\item For the set of fixed points given by the solution of 
\begin{eqnarray}
\beta x_0(g-1)-\gamma ln(g)-\beta y_0=0
\end{eqnarray}
the factor $(-\gamma+x_0\beta g^*)$ decides the fate of the pandemic/epidemic: (i) if $g^*\rangle\frac{\gamma}{\beta x_0}$, $g^*$ corresponds to unstable fixed point, the condition which the pandemic/epidemic persists, whereas, (ii) if $g^*\langle\frac{\gamma}{\beta x_0}$, $g^*$ becomes stable fixed point, the condition at which the pandemic/epidemic moves to endemic.
\end{itemize}
}
\noindent\textbf{Proof:} \textit{Fixed points of the equation \eqref{geq} can be obtained by putting $\frac{dg}{dt}=0$ and solving for $g\rightarrow g^*$. The fixed points are $g^*=0$, and solutions of $g^*$ of
\begin{eqnarray}
\label{fix}
\beta x_0(g-1)-\gamma ln(g)-\beta y_0=0
\end{eqnarray}
The fixed point $g^*=e^{-\frac{\beta}{\gamma}z^*}=0$ is the condition when $z^*\rightarrow\infty$ which can happen at significantly long time ($t\rightarrow\infty$). This is the condition at which all the infected people in a certain demographic regime become recovered. This means $I\rightarrow R,~R\rightarrow N$, where, $N$ is the exact population in the demographic regime. This is, in fact, the condition of \textit{endemic} i.e. the pandemic/epidemic ends.} \\

{\noindent}Linear stability analysis of the equation \eqref{geq} is done in order to understand the behavior of epidemiological dynamics near the fixed point by expanding $g$ around the fixed point $g^*$: $g=g^*+w$, where, $w$ is an arbitrary variable $w\langle 1$. Putting this expression to equation \eqref{geq} and expanding $F(g^*+w)$ around $g^*$ using Taylor's series expansion keeping only linear terms, we have,
\begin{eqnarray}
\frac{dw}{dt}=F(g^*)+w\left. \frac{dF}{dg}\right|_{g\rightarrow g^*}+O(w^2)\sim w\left[-\gamma +\beta x_0g^*\right]
\end{eqnarray}
such that, one can obtain: $w(t)\sim w_0e^{[-\gamma+x_0\beta g^*]t}$, and one can get the approximate solution given by: $g(t)\sim g^*+w_0e^{[-\gamma+x_0\beta g^*]t}$. For the fixed point $g^*=0$, $g(t)\sim g^*+w_0e^{-\gamma t}$, where, eigenvalue $\lambda=-\gamma\langle 0$ indicating stable fixed point such that $\displaystyle\lim_{t\rightarrow\infty} g(t)\rightarrow g^*=0$. Since $g^*=e^{\frac{\beta}{\gamma}z^*}$, $g^*=0$ only when $z^*\rightarrow\infty$ as $t\rightarrow\infty$ which is the condition: $I\rightarrow R$, $R\rightarrow N$ at $t\rightarrow\infty$. \\

{\noindent}The set of fixed points provided by equation \eqref{fix}, the approximate solution is given by $g(t)\sim g^*+w_0e^{[-\gamma+x_0\beta g^*]t}$ with $s^*\ne 0$. The following two conditions in the stability analysis can be observed:
\begin{itemize}
\item If $-\gamma+x_0\beta g^*\langle 0$, such that, $g^*\langle\frac{\gamma}{\beta x_0}$, then $g^*$s are stable fixed point, where, $\displaystyle\lim_{t\rightarrow\infty}g(t)\rightarrow g^*\ne 0$. These fixed points may correspond to metastable states, where, pandemic/epidemic is controlled as $z^*\rightarrow large$, in the case of multiple epidemic/pandemic waves similar to Covid-19 pandemic.
\item If $-\gamma+x_0\beta g^*\rangle 0$, where, $g^*\rangle\frac{\gamma}{\beta x_0}$, then $g^*$s are unstable fixed point, where, $\displaystyle\lim_{t\rightarrow\infty}g(t)\rightarrow\infty$ (blows up). The condition could correspond to the state at which pandemic/epidemic is active.
\item The case $-\gamma+x_0\beta g^*= 0$, with $g^*=\frac{\gamma}{\beta x_0}$, then $g^*$s is saddle fixed point, where, $\displaystyle\lim_{t\rightarrow\infty}g(t)\rightarrow g^*=constant$.
\end{itemize}

\noindent\textbf{Lemma 2:} \textit{The equivalent ODE \eqref{geq} of SIR model exhibit asymptotic stability bifurcation with the bifurcating parameter $g^*=\frac{\gamma}{\beta x_0}$. 
}\\

\noindent\textbf{Proof:} \textit{Since the dynamics described by the equation \eqref{geq} shows stable for $g^*\langle\frac{\gamma}{\beta x_0}$, and unstable for $g^*\rangle\frac{\gamma}{\beta x_0}$, it clearly indicates that the dynamics of $g(t)$ exhibits asymptotic stability bifurcation \cite{strogatz} as shown in Fig. 1. The bifurcation diagram indicates that when $g^*\langle\frac{\gamma}{\beta x_0}$ there could be two possibilities either there is no pandemic/epidemic or if there is pandemic/epidemic then it is the case of endemic. Since $g^*=e^{-\frac{\beta}{\gamma}z^*}$, we have in this case $z^*\rangle\frac{\gamma}{\beta x_0}ln\left[\frac{\beta x_0}{\gamma}\right]$. For a fixed initial condition $x_0$ we have $\gamma\rangle\rangle\beta$ because of fast recovering as compared to quite slow infection rate in this situation and so one can approximate $ln\left[\frac{\beta x_0}{\gamma}\right]\sim\left[\frac{\beta x_0}{\gamma}-1\right]$. In this state, we have, $z^*\rangle x_0-\frac{\gamma}{\beta}$. Hence, when $x_0\sim\frac{\gamma}{\beta}$ the epidemic ends and this condition depends on the parameters $\gamma$ and $\beta$ sensitively. As $g^*$ departed from $g^*=0$ ($g^*\rangle 0$), pandemic/epidemic starts but it is under controlled till $g^*\langle\frac{\gamma}{\beta x_0}$. Once it reach $g^*=\frac{\gamma}{\beta x_0}$, the pandemic/epidemic becomes active and as $g^*\rangle\frac{\gamma}{\beta x_0}$, the pandemic/epidemic becomes uncontrollable. Hence, we need to manage the parameter $\gamma$ and $\beta$ so that $g^*\langle\frac{\gamma}{\beta x_0}$ to become controllable.}\\

{\noindent}The analysis of the equivalent SIR epidemiological model clearly shows that epidemiological dynamics and its controllability are quite dependent on the parameters $\beta$ and $\gamma$ and are very sensitive to these parameters. Further, these parameters becomes quite sensitive variables not constants which could be dependent on various factors, namely, pandemic/epidemiological features, the nature of viral dynamics as in the case of ongoing Covid-19 pandemic (viral virulence, health index, temperature, pressure etc) \cite{Singh, Mandal} etc. Hence, it is important to search for the dynamics of these parameters and correlate with controlling factors of the pandemic/epidemic triggering the changes in these parameters such as social distance, lockdown etc \cite{Chanu}.\\

\begin{figure}[ht]
\includegraphics[height=8cm,width=11cm]{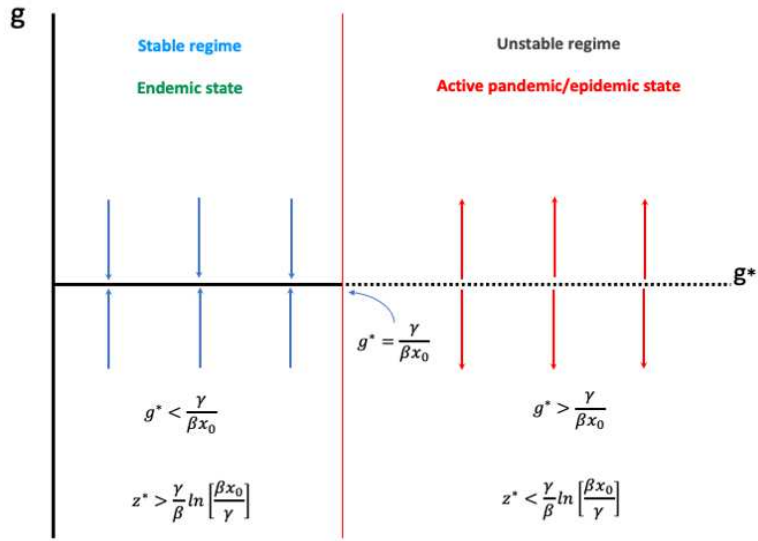}
\caption{Bifurcation diagram of the equivalent SIR model equation.}
\end{figure}

%%%%%%%%%%%%%%%%%%%%%%%%%%%%%%%%%%%%%%%%%%%%%%%%%%%%%%%%%%%%%%%%%%%%%%%%%%%%%%%%%%%%%%%%%%

\noindent\textbf{\large Stochastic description of the SIR model}\\\\
{\noindent}The SIR model described by equation \eqref{sir} can be translated into two reactions model: $X+Y\stackrel{\beta}{\rightarrow}2Y;~Y\stackrel{\gamma}{\rightarrow}Z$ \cite{Kermack,Pastor}. Consider the activities of these reactions are random in nature and obey Markov process \cite{McQuarrie}. Then, the state of the system at any instant of time "$t$" can be defined by a state vector, $[x,y,z]^T=S(t);~x,y,z\in I(integer)$ represent populations of susceptible, infectious and recovered in the demographic region, where, $T$ is the vector transpose. The transition of the state vector during time interval $[t,t+\Delta t]$ can be engineered as the birth or death of one molecule, and $\left[\begin{matrix}x\\y\\z \end{matrix}\right]\rightarrow\left[\begin{matrix}x\pm 1\\y\pm 1\\z\pm 1\end{matrix}\right]$ are the possible transition states in the model during this time interval. Further, the relationship between stochastic rate constant $\{c_j\};j=1,2$ and classical rate constants $\beta$ and $\gamma$ is given by the relation $c_1=\beta V^{1-\nu}=\frac{\beta}{V}$ and $c_2=\gamma V^{1-\nu}=\gamma$, where, $\nu$ is the stoichiometric ratio and $V$ is the system size \cite{Gillespie}. The probability of the state $\left[\begin{matrix}x\\y\\z \end{matrix}\right]$ at time $(t+\Delta t)$ is obtained from the detailed balance condition of loss and gain contributions from the reactions, and can be obtained the Master equation of the model \cite{McQuarrie}. The Master equation thus obtained is described as,
\begin{eqnarray}
\label{xy1}
\frac{\partial P(x,y,z; t)}{\partial t}  &=& \frac{c_1}{V} (x+1)(y-1) P(x + 1,y-1,z; t) + c_2 (y+1) P(x,y+1,z-1; t)\nonumber\\
&& - \left[\frac{c_1}{V} xy + c_2 y\right]P(x,y,z; t)
\end{eqnarray}
We know that $x+y+z=N$ such that $z=N-x-y$. This relation allows us dimensional reduction from three dimension to two dimension in the Master equation \eqref{xy1}. Now we can write equation \eqref{xy1} as following by reducing to the state $S(t)=[x,y,z]^T\rightarrow [x,y]^T$, 
\begin{eqnarray}
\label{xy2}
\frac{\partial P(x,y; t)}{\partial t} =  \frac{c_1}{V} (x+1)(y-1) P(x + 1,y-1; t) + c_2 (y+1) P(x,y+1; t) - \left[\frac{c_1}{V} xy + c_2 y\right]P(x,y;t)
\end{eqnarray}
This Master equation \eqref{xy2} can be solved by using generating function technique. For this we define a generating function in two variable given by,
\begin{eqnarray}
\label{xy3}
G(s,r;t)=\sum\limits_x \sum\limits_y s^{x}r^{y}P(x,y;t)
\end{eqnarray} 
The boundary condition to be taken is $G(r,s;0)=r^Ns^N$, and normalization condition to the generating function. Now multiply \eqref{xy2} by $s^x r^y$ in both sides, then doing $\sum_x \sum_y$ and after simplifying algebra, we have,
\begin{eqnarray}
\label{xy4}
\frac{\partial G(s,r; t)}{\partial t} =  \frac{c_1}{V} r(r-s) \frac{\partial^{2}G(s,r; t)}{\partial s \partial r} + c_2 (1-r) \frac{\partial G(s,r;t)}{\partial r}
\end{eqnarray}
This spatiotemporal partial differential equation \eqref{xy4} can be solved using Lagrange-Charpit characteristics method \cite{Delgado}. Using this method to the equation \eqref{xy4}, we have,
\begin{eqnarray}
\label{xy5}
\frac{dt}{1} =  - \frac{ds}{\frac{c_1}{V} r(s-r)\frac{\partial G}{\partial r}} = - \frac{dr}{c_2 (r-1)}
\end{eqnarray}
{\noindent}From first and third part of equation \eqref{xy5}, we have, $\frac{dt}{1} = -\frac{dr}{c_2 (r-1)}$, and then integrating it we have the following solution, 
\begin{eqnarray}
\label{xy6}
(r-1)e^{-c_2 t} = A(constant)
\end{eqnarray}
From the second and third part of equation \eqref{xy4}, we have the following equation,
\begin{eqnarray}
\label{xy7}
\frac{\partial G}{c_2 (r-1)} = -\frac{c_2 V}{c_1}\frac{r-1}{r}\frac{1}{r-s}ds
\end{eqnarray}
Integrating equation \eqref{xy7}, we get,
\begin{eqnarray}
\label{xy8}
G(s,r;t) = \frac{c_2 V}{c_1}\frac{r-1}{r}ln(r-s) B\\
\end{eqnarray}
where, $B$ is another constant of integration. From equation \eqref{xy7} and \eqref{xy8}, we can write the expression for $G(r,s;t)$ through the constants $A$ and $B$ as the following,
\begin{eqnarray}
\label{xy9}
G(s,r;t) = \frac{c_2 V}{c_1}\frac{r-1}{r}ln(r-s) f[(r-1)e^{-c_2 t}]
\end{eqnarray}
The functional form of $f$ in equation \eqref{xy9} can be calculated using the boundary condition in $G(s,r;t)$ i.e. $G(s,r;0)=s^Nr^N$ and putting it to the equation \eqref{xy9}. Further, taking the term $ln(r-s)\rightarrow ln|r-s|$ for the sake of consideration of the magnitude of the difference $|r-s|$ or $|s-r|$. One reason for taking this assumption is that this difference is generally small during the active pandemic/epidemic state, and at the beginning of the pandemic/epidemic $ln|r-s|\rightarrow ln|s|$, whereas, at the endemic state $ln|r-s|\rightarrow ln|r|$. Considering this approximation, one can write: $ln|r-s|=\displaystyle\sum_{n=1}^{\infty}(-1)^{n-1}\frac{\left[|r-s|-1\right]^n}{n}$. Now after applying the boundary condition, replacing the argument $r$ in the function form $f(r)$ by $r\rightarrow (r-1)e^{-c_2t}$ and putting this power series expansion to the generating function in equation \eqref{xy9}, we have, 
\begin{eqnarray}
\label{xy10}
G(s,r;t) = \frac{1}{r}e^{c_2t}\left[(r-1)e^{-c_2t}+1\right]^{N+1}s^N
\frac{\sum\limits_{n=1}^{\infty} (-1)^{n-1}\frac{(|r-s|-1)^n}{n}}{\sum\limits_{n=1}^{\infty} (-1)^{n-1}\frac{\left|(r-1)e^{-c_2t}-s\right|^n}{n}}
\end{eqnarray}
We are interested in the dynamics of mean infected population $\langle y\rangle$ because if we know this dynamics, other dynamics $\langle x\rangle$ and $\langle z\rangle$ can be correlated and can be obtained in the similar manner of calculating $\langle y\rangle$ and using $x+y+z=N$. Hence, we show only the calculation and analysis of $\langle y\rangle$ which can be done using the generating function $G(s,r;t)$ through the relation: $\displaystyle\langle y \rangle = \frac{\partial G(s,r;t)}{\partial r}|_{r=1,s=1}$. Now, using the expression of $G(s,r;t)$ in equation \eqref{xy10} and using the relation mentioned, we can able to reach the following expression of mean infectious population after some algebra,
\begin{eqnarray}
\label{xy11}
\langle y \rangle = N+1-e^{c_2 t}+(e^{c_2 t}-1)\frac{\sum\limits_{n=1}^{\infty}(-1)^{2(n-1)}}{\sum\limits^{\infty}_{n=1}(-1)^{2n-1}\frac{1}{n}} 
\end{eqnarray}
We know that in the expression in the numerator of equation \eqref{xy11} $\displaystyle\sum_{n=1}^{\infty}(-1)^{2(n-1)}\rightarrow\sum_{n=1}^{\infty}1=1+1+...+\infty=\frac{1}{2}$ because $2(n-1)$ is always positive for any integer value of $n$ and the sum can be interpreted by zeta function regularization of the Riemann zeta function. Similarly, the denominator of this equation \eqref{xy11} can be written as, $\displaystyle\sum^{\infty}_{n=1}(-1)^{2n-1}\frac{1}{n}=\sum_{n=1}^{\infty}\frac{1}{n}$. This sum expression is the harmonic series, and the partial sums of the series have logarithmic growth, and hence can be written as: $\displaystyle\sum^{k}_{n=1}\frac{1}{n} = ln(k)+\gamma + \epsilon_k \leq ln(k)+1$, where, $\gamma$ is the Euler-Mascheroni constant and $\epsilon_k \sim \frac{1}{2k}$ which approaches $0$ as $k$ goes to infinity. Putting these expressions to the equation \eqref{xy11} and $c_2=\gamma$, we can able to reach in the following simplified form of the $\langle y\rangle$
\begin{eqnarray}
\label{xy12}
\langle y \rangle = N-\frac{ln(k)}{1 + ln(k)}\left(e^{\gamma t}-1\right) 
\end{eqnarray}
Now we can have the following analysis.\\

\noindent\textbf{Theorem 1.} \textit{The mean infected population in a pandemic/epidemic,
$\langle y \rangle$ has the following properties:
\begin{itemize}
\item If the recovering rate is very small, then: $\displaystyle\lim_{\gamma\rightarrow 0}\langle y \rangle=N$ (mean infected population will become initial infected population).
\item Endemic time is given by, $t_E=\frac{1}{\gamma}\ln\left[N\left(1+\frac{1}{\ln[k]}\right)+1\right]$.
\end{itemize}}

{\noindent}\textbf{Proof:} \textit{From the equation (\ref{xy12}), if we take small $\gamma$ limit then one can get, $\displaystyle\lim_{\gamma\rightarrow 0}\langle y\rangle=N$. This means that if the recovering rate is very small at any instant of time $t$, then mean infected population in the demographic region does decreased but remain stagnant to the mean infected population at that time "$t$". Hence, there should be some policy to increase $\gamma$ in order to have $\langle y\rangle\rightarrow 0$.}\\

\noindent\textit{The endemic is the condition when $\langle y\rangle\rightarrow 0$. The time taken to reach this condition can be defined as the endemic time $t_E$ and can be calculated by taking the limit $\langle y\rangle\rightarrow 0$. Applying this limit to the equation \eqref{xy12}, we get the endemic time given by, $t_E=\frac{1}{\gamma}\ln\left[N\left(1+\frac{1}{\ln[k]}\right)+1\right]$. Further, for large values of $k$, $ln(k)\rightarrow large$, such that, $\displaystyle\lim_{k\rightarrow\infty}ln(k)\rightarrow\infty$. In this limit, $\displaystyle t_E\approx\frac{1}{\gamma}ln(N+1)$. This result indicates that $t_E\propto\frac{1}{\gamma}$ i.e. larger the recovering rate quicker the pandemic/epidemic ends.
}
\\

{\noindent}Now, we study the effect of fluctuations or noise in the stochastic SIR model system and how does it drives the epidemiological dynamics at various dynamical states. One way to analyze this fluctuations characteristics is to measure \textit{Fano factor} which is given by $F=\displaystyle\frac{\sigma_y^2}{\langle y\rangle}$ \cite{Fano}, where, $\sigma_y$ is the standard deviation with respect to the variable $y$. The dynamics of the system can be characterized by fluctuations as follows: (i) if $F\langle 1$, such that  the variance ($\sigma_y^2< \langle y\rangle$) then the dynamics is \textit{sub-Poissonian} process in which the dynamical process suppresses fluctuations; (ii) if $F=1$, where, $\sigma_{y}^{2}= \langle y\rangle$, indicates that the process is \textit{Poissonian} in which the processes in the system's dynamics become statistically independent of each other; and (iii) if $F\rangle 1$, where, $\sigma_y^2> \langle y\rangle$ which indicates \textit{super-Poissonian or noise enhancement} process, where, the role of noise in regulating system's dynamics becomes significant \cite{Zou,Davidovich,Chanu}. Now, to obtain the expression for $F$, we calculated variance $\sigma_y^2$ using the expression for $G(s,r;t)$ in equation \eqref{xy10} as in the following way,
\begin{eqnarray}
\label{xy13}
\sigma_y^2 &=&  \left.\frac{\partial^2G}{\partial r^2}\right|_{s=1,r=1}+\left.\frac{\partial G}{\partial r}\right|_{s=1,r=1}-\left[\left.\frac{\partial G}{\partial r}\right|_{s=1,r=1}\right]^2\nonumber\\
&=&\frac{1}{\bigg[\sum\limits_{n=1}^{\infty} \frac{(-1)^{2n-1}}{n}\bigg]^2}\Bigg\{ \sum\limits_{n=1}^{\infty}\frac{(-1)^{2n-1}}{n} \Bigg[ [N(N+1)e^{-c_2 t}]\sum\limits_{n=1}^{\infty} \frac{(-1)^{2n-1}}{n} + 2N\sum\limits_{n=1}^{\infty}(-1)^{2(n-1)}\nonumber\\
&&+ (e^{c_2 t}-e^{-c_2 t})\sum\limits_{n=1}^{\infty} (-1)^{2n-3}(n-1) \Bigg] -\Bigg[(N+1-e^{c_2 t})\Bigg(\sum\limits_{n=1}^{\infty}\frac{(-1)^{2n-1}}{n} \Bigg)+(e^{c_2 t}-1)\sum\limits_{n=1}^{\infty}(-1)^{2(n-1)}\Bigg]\nonumber\\
&& \times \Bigg[(N+2-e^{c_2 t})\Bigg(\sum\limits_{n=1}^{\infty}\frac{(-1)^{2n-1}}{n} \Bigg)+(2e^{-c_2 t}+e^{c_2 t}-1)\sum\limits_{n=1}^{\infty}(-1)^{2(n-1)}\Bigg]\Bigg\}
\end{eqnarray}
Substituting the equation \eqref{xy13} and \eqref{xy12} to the expression for Fano factor and after some algebra and simplification, we get the following expression for $F$,
\begin{eqnarray}
\label{xy14}
F &=& N(N+1)\frac{e^{-\gamma t}}{\langle y \rangle}-\langle y \rangle -1-\frac{e^{-\gamma t}}{ln(k)+1} - \frac{N - \frac{e^{\gamma t} -e^{-\gamma t}}{12}}{(ln(k)+1)\langle y \rangle}
\end{eqnarray}
Further, for the case large $k$ limit, where, $\displaystyle\lim_{k\rightarrow\infty}ln(k)\rightarrow\infty$, the above equation \eqref{xy14} becomes, $\displaystyle F=N(N+1)\frac{e^{-\gamma t}}{\langle y \rangle}-\langle y \rangle -1$.\\

\noindent\textbf{Theorem 2.} \textit{The Fano factor analysis for the stochastic SIR pandemic/epidemic provide us the following observations:
\begin{itemize}
\item Noise enhanced process provide us estimation of pandemic/epidemic time with respect to endemic time $t_E$ by,
\begin{eqnarray}
\label{xy15}
2t_E-t>\frac{1}{\gamma}ln\left[\langle y\rangle(\langle y\rangle+2)\right]
\end{eqnarray}
\item For large $\langle y\rangle$ large limit, $t_E-\displaystyle\frac{1}{2}t>\frac{1}{\gamma}ln\left[\langle y\rangle\right]$.
\item For the small $\langle y\rangle$ limit, $t_E-\displaystyle\frac{1}{2}t>\frac{1}{2\gamma}ln\left[\frac{3}{2}\langle y\rangle+ln(2)-1\right]$, with the condition that $\langle y\rangle >\frac{2}{3}[1-ln(2)]$.
\end{itemize}
}

\noindent\textbf{Proof:} \textit{From the equation \eqref{xy14} the condition for the noise enhanced or super-Poissonian process can be obtained by putting the condition $F>1$. In this situation, one can get,}
\begin{eqnarray}
\label{xy16}
N(N+1)\frac{e^{-\gamma t}}{\langle y \rangle}>\langle y \rangle +2+\frac{e^{-\gamma t}}{ln(k)+1} - \frac{N - \frac{e^{\gamma t}-e^{-\gamma t}}{12}}{(ln(k)+1)\langle y\rangle} 
\end{eqnarray}
\textit{Now, for the large $k$ limit, the last two terms can be neglected as they are very small in this limit. Further, in this limit, we have, $ln[N(N+1)]=ln(N)+ln(N+1)\approx\gamma 2\gamma t_E$ as $t_E=\frac{1}{\gamma}ln(N+1)$. Then putting these expressions and rearranging the terms, one can reach the equation \eqref{xy15}. From the relation one can estimate approximate time interval to reach endemic $(2t_E-t)$ which depends on $\gamma$ for a fixed value of $\langle y\rangle$ at a particular instant of time $t$.}\\

\noindent\textit{The straight forward proof can be done from the equation \eqref{xy15} by taking large $\langle y \rangle$ limit, where, $\langle y \rangle[\langle y \rangle+2]\sim\langle y \rangle^2$. Rearranging the terms one can prove it. Similarly, for small $\langle y \rangle$ limit, one can take $ln\langle y \rangle\sim\langle y \rangle-1+o(\langle y \rangle^2)$, and $ln(1+\langle y \rangle)\sim\langle y \rangle+o(\langle y \rangle^2)$. Putting these expressions to the equation \eqref{xy15} and rearranging the terms one can prove the relation. In order to have super-Poissonian process one has to have $\langle y\rangle >\frac{2}{3}[1-ln(2)]$.}\\

\noindent\textbf{Noise can modulate the fate of the epidemic}\\
{\noindent}We study the impact of noise (intrinsic) in the model by calculated Fano factor given in equation \eqref{xy14}. The variation of the Fano factor $F$, which can characterize the noise associated with the system's dynamics, as a function of $t$ and $\gamma$ in the model are shown in Fig. 4. From the figure one can easily observe that at the early stage of the pandemic/epidemic ($t<3$), the epidemic process is sub-Poissonian where the noise is suppressed in the dynamical behavior of the system though $F$ rapidly increases with $t$ (Fig 4(A)). At around $t\sim 3$, the dynamical processes in the system become statistically uncorrelated (Poissonian process) to each other due to the pandemic/epidemic becomes active. If $t>3$, as time progress $F>1$, the process become super-Poissonian or noise enhanced process which is dissipative and the noise become one important parameter driving the system to active pandemic situation.\\

{\noindent}Next, in Fig 4(B), we tried to understand the behavior of $F$ as a function of ($\gamma$) $(\gamma=[0 \ to \ 1])$ and found that the behavior of $F$ is almost the same for as (Fig. 4(A)).  Here also, we could see that $F<1$ for small values of $\gamma$ (sub-Poissonian process). As $\gamma$ increases $F\rightarrow 1$ (Poissonian process) and $F>1$ as $\gamma\rightarrow large$ (noise-enhanced process).
{\noindent}We thus observe that noise are integral parts of SIR model dynamics. These parameters can be a key factor for the system.\\

\begin{figure}[ht]
\includegraphics[height=5cm,width=14cm]{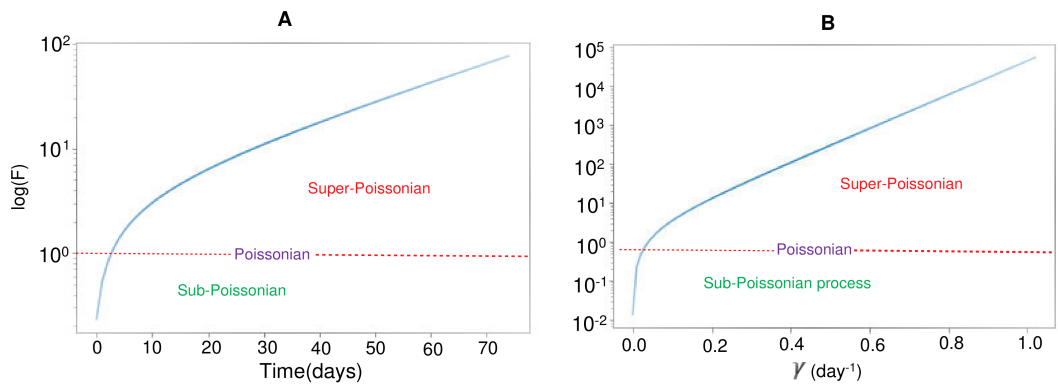}
\caption{\textbf{(A)} Variation of Fano factor (calculated from Master equation) with respect to time $t$. \textbf{(B)} Variation of Fano factor (calculated from Stochastic Master equation) with respect to  time $\gamma$. }
\end{figure}
%%%%%%%%%%%%%%%%%%%%%%%%%%%%%%%%%%%%%%%%%%%%%%%%%%%%%%%%%%%%%%%%%%%%%%%%%%%%%%%%%%%%%%
\vskip 0.3cm
{\noindent}\textbf{\large Method for parameter estimation and its dynamics}\\ 
\\
{\noindent}The estimation of epidemiological parameters ($\beta$, $\gamma$) in SIR model can be done using approximation Bayesian computation (ABC) based on stochastic simulation algorithm (SSA) due to Gillespie algorithm which is based on likelihood free inference methods and developed for stochastic systems \cite{Toni}. This method is implemented on the dynamics of biochemical model \cite{Lie} of SIR model to evaluate marginal posterior distributions to estimate values of the parameters for different situations in the current Covid-19 pandemic with reference to Indian data. In the parameter estimation, the likelihood calculations are replaced by the simulation of the model system within certain period of time. This method ABC SSA, which is one of the ABC SMC (ABC sequential Monte carlo) methods, provides better parameter estimation than the existing ABC algorithms \cite{Toni}.\\
 
{\noindent}The ABC SSA method can be explained as follows \cite{Toni}. Consider $x$ be a given data. Then from this data, for a desired sample size, $N$, one can define an acceptance threshold, $\epsilon$. Let $\theta$ be a parameter vector which we have to estimate using ABC SSA. Given the prior distribution $\pi(\theta)$, the observed data be denoted as $x^*$. The parameter values are generated through a distributions, $\pi(\theta |d(x,x^*) \leq \epsilon)$ \cite{Toni}.\\
While $i<N$ 
\begin{itemize}
\item Sample $\theta^*$ from the prior distribution $\pi(\theta)$
\item Simulate a model data set $x*$ from $\pi(x|\theta^*)$
\item If $d(x,x^*) \leq \epsilon$, accept $\theta^*$ into the posterior distribution
\item $i = i +1$
\end{itemize} 
After estimating the parameter values, the SSA which is Gillespie algorithm is implemented in the simulation. The reaction network of the SIR model is simulated using SSA \cite{Gillespie} to understand the dynamics of pandemic/epidemic. The Gillespie algorithm is based on two random processes, which reaction will fire at what time \cite{Gillespie}. The algorithm consist two random variables $r_1$ and $r_2$ such that reaction time is computed using $\tau=-\frac{1}{a_0}ln(r_1)$, where, $a_0\sum_i a_i$,$a_i$ is the $ith$ propensity function and $kth$ reaction will fire when it satisfy, $\displaystyle\sum^{k}_{i=1} a_i \leq a_0 r_2< \sum^{k+1}_{i=1}a_i$. 

%%%%%%%%%%%%%%%%%%%%%%%%%%%%%%%%%%%%%%%%%%%%%%%%%%%%%%%%%%%%%%%%%%%%%%%%%%%%%%%%%%%%%%%%%

\vskip 0.7cm
\noindent\textbf{Parameter estimation of SIR model and sensitivity analysis}\\
{\noindent}We used the equivalent two reactions set \cite{Pastor} of the classic SIR model \cite{Kermack} and applied ABC SSA algorithm as stated in the method above to estimate parameters $\beta$ and $\gamma$. In order to understand the sensitivity of these parameters on the current Covid-19 pandemic dynamics and the importance of government policies to control this ongoing pandemic, we considered \textit{lockdown} as government policy and dynamics of infected population during the \textit{lockdown} period to estimate the model parameters. The \textit{lockdown-1} had 21 days duration (Fig. 2 left panel) and used this data to estimate the parameters using ABC SSA algorithm. The posterior distributions of $\beta$ and $\gamma$ (Fig. 2 middle and right panels) allow us to estimate the values of the model parameters as given in the Table 1. These parameter values are used to simulate the SIR reaction network to get the infected population dynamics $I$ or $y$ (Fig. 2 left panel blue line) to compare with the real data and found that stochastically simulated model prediction is in good agreement with the real data (real data considered from WHO represented by the red dots for t (days) = [1,5,10,15,20]).  Here the curve in the figure (Fig. 2 left panel) represents the pointwise median predictions of the stochastic SIR model for the infected population, and the shaded portion represents the  $95\%$ credible regions.\\

\begin{figure}[ht]
\includegraphics[height=15cm,width=13cm]{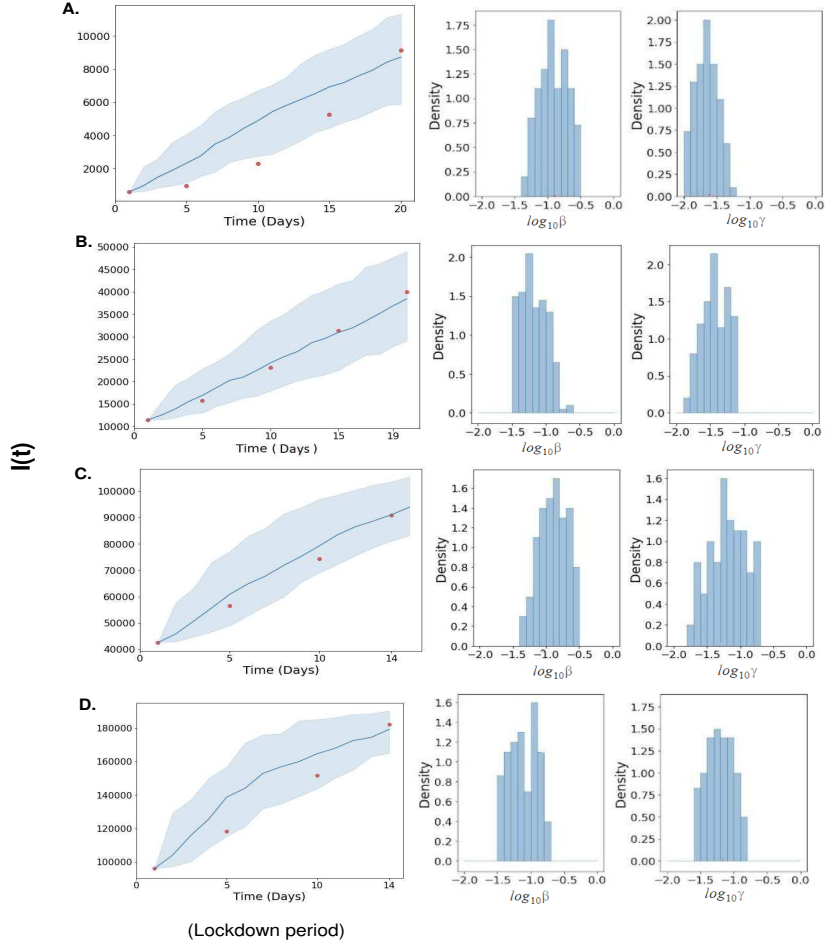}
\caption{\textbf{A. Lockdown 1:} The dynamics of the infected population(left) using estimated parameters by ABC SSA algorithm applied to SIR reaction set. The shaded portion indicates $95\%$ credible region. The blue curve is the simulated dynamics of $I(t)$ using SSA and estimated parameter values. Red points are the real data values for days 1,5,10,15,20 during lockdown period. The posterior histogram of $\beta$ and $\gamma$ value from which the values of the parameters are estimated are shown in middle and right panels. Similar plots of \textbf{B. Lockdown 2:} (lockdown period taken is 19 days, real data points taken for comparison are 1,5,10,15,20 days), \textbf{C. Lockdown 3:} (lockdown period taken is 14 days, real data points taken for comparison are 1,5,10,14 days) and \textbf{D. Lockdown 4:} (lockdown period taken is 14 days, real data points taken for comparison are 1,5,10,14 days) are shown in the panels B, C and D respectively.}
\end{figure}

{\noindent}Similar plots of \textbf{lockdown 2}, \textbf{lockdown 3} and \textbf{lockdown 4} are shown in the panels B, C and D respectively in Fig. 2. From the results show clearly that the posterior distributions of $\beta$ and $\gamma$ are quite different and the estimated values of these parameters significant variations as shown in Table 1 for different lockdown periods. Simulated results indicating the dynamics of the infected population $I(t)$ using the estimated parameter values are in good agreement with the real data points taken from WHO approved Covid-19 dashboard. This behavior of the estimated parameters indicates that these parameters are quite sensitive to the changes in the dynamics of the infected population and should not be taken constant as has been taken in most of the simulations of SIR model. \\

\begin{table*}[ht]
\begin{center}
{\bf Table 1: List of estimated parameters} \\
%\begin{adjustbox}{width={\textwidth},totalheight={\textheight},keepaspectratio}% 
\begin{tabular}{|l|p{2.5cm}|p{2.5cm}|p{1.5cm}|p{1.5cm}|p{2.5cm}|p{1.5cm}|}
 \hline %\multicolumn{5}{}{} \\ \hline
\bf{}&\bf{Start date}&\bf{End date}& \bf{$\beta$}&  \bf{$\gamma$} &\bf{Reproduction number}	\\ \hline
\textbf{Lockdown 1}&25th March 2020&14th April 2020&$0.122265$&$0.020753$&$5.891404$  \\ \hline
\textbf{Lockdown 2}&15th April 2020&3rd May 2020&$0.082674$&$0.038415$&$2.152128$	\\ \hline
\textbf{Lockdown 3}&4th May 2020&17th May 2020&$0.087931$&$0.046105$&$1.907193$  \\ \hline
\textbf{Lockdown 4}&18th May 2020&31st May 2020&$0.078204$&$0.056576$&$1.382283$  \\ \hline
\end{tabular}
%\end{adjustbox}%
\end{center}
\end{table*}

\vskip 0.5cm
\noindent\textbf{Effectiveness of control strategies can be captured in the parameter dynamics}\\
{\noindent}The results of estimated parameters of SIR model show that continuous implementation of lockdowns allows to decrease in the reproduction number indicating lockdown strategy could be one of the potential strategies which can control this complicated Covid-19 pandemic (Fig. 3). Since, this $R_0$ values are dependent on the parameter values of $\beta$ and $\gamma$, the dynamics of these parameters, which can be estimated using ABC SSA algorithm, can be used as the forecasting parameter of pandemic states as well as effectiveness of any strategy implemented to control this pandemic. The impact of consecutive lockdown characterized by $l$ on the parameters $\beta$ and $R_0$ decreased with power law: $\beta\sim l^{-0.325}$ and $R_0\sim l^{-1.156}$. However, since the value of $R_0>1$ till $l=4$ the pandemic is not under control. Since $R_0$ with $l$ follow the relationship $R_0\sim l^{-1.156}$ (exact form is $R_0=5.8\times l^{-1.156}$), the pandemic can be controlled ($R_0<1$) approximately after $l\ge 5$. During the active Covid-19 pandemic state $\gamma$ and $\frac{\gamma}{\beta}$ follows power growth nature: $\gamma\sim l^{0.669}$ and $\frac{\gamma}{\beta}\sim l^{0.872}$ (Fig. 3).\\

{\noindent}From Mandelbrot's fractal theory \cite{Mandelbrot}, if a certain system parameter follows power law behavior with a scale parameter, it is the indicator of fractal nature of the system with respect to that scale parameter. This fractal nature of the system is the signature of self-organization \cite{Kaufmann}. Since the behavior of the SIR parameters follow power law nature with lockdown parameter $l$, the impact of lockdown to people in a certain demographic region is to trigger more organized behavior with lockdown parameter $l$ reflected in the pandemic dynamics. The more organized faster is to control the pandemic with respect to lockdown. 

\begin{figure}[ht]
\includegraphics[height=11cm,width=11cm]{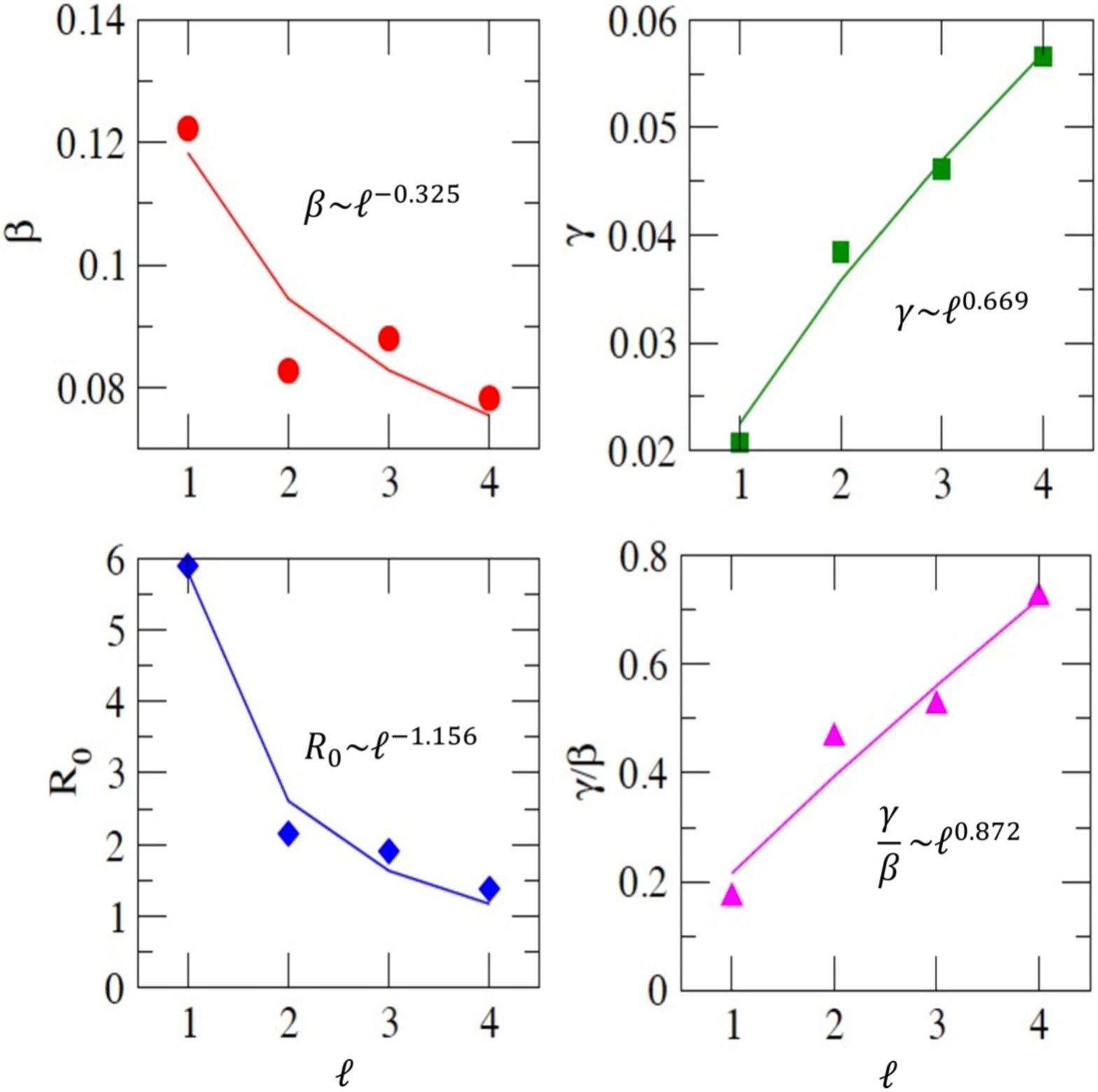}
\caption{Dependence of SIR on lockdown parameter $l$. The solid lines are the fitted lines on the data points with equations given in each panel. The goodness of the fit characterized by $r^2-value$ ranges from 0.9135-0.9567.}
\end{figure}

\vskip 0.7cm
{\noindent}\textbf{\large Conclusion}\\
\\
{\noindent}The ongoing Covid-19 pandemic is a dreaded threat to the global human population and India being second most affected in the world so far. The study of the classic epidemiological SIR model fitting to the current pandemic to search for strategies to control this pandemic leading to endemic. The deterministic analysis of the exact analytical explicit solution of the SIR model indicates that the calculated fixed points correspond to equilibrium states of the pandemic from which control strategies can be formulated. Further, these fixed points are also quite sensitive to the epidemiological parameters $\frac{\gamma}{\beta}$ giving us the possibility of designing control strategies. For example, $\beta$, which is infection rate, can be related to various strategies, namely, lockdown, social distancing, isolation, etc to decrease $\beta$ \cite{Chanu}. On the other hand, to increase $\gamma$ the strategies such as vaccination, exercise, increase immunity, etc can be correlated. Even the parameter $\frac{\gamma}{\beta}$ can be used as the indicator of whether a certain pandemic/epidemic is going to be active or endemic. Similar results were also found in the case of stochastic analytical results. The additional information we got from the stochastic analysis is the role of noise in triggering pandemic/epidemic. The origin of noise in the case of epidemiological study could be various factors, for example, random movement of the people without following WHO or ICMR rule and regulation, human diffusion from one region to another, people who are against the government policies/strategies, etc which could trigger pandemic more active. Hence, these noises needed to be controlled properly to control the pandemic.\\ 

{\noindent}From the study of the analytical results (both deterministic and stochastic) of the SIR model, clearly show the sensitive dependence of the epidemiological parameters on the epidemiological dynamics and states. We, therefore, estimated the values of epidemiological parameters using the designed ABC SSA algorithm taking the Indian government's lockdown strategy as the factor causing changes in the epidemiological dynamics and states. The estimated values of the parameters are used to simulate the model system and found to be in good agreement with the real data. The estimated values of these parameters are found to be changed and sensitive to the lockdown strategy. In India's case, the lockdown was found to be the more effective strategy implemented by the Indian government to reduce infection overpopulation. The results show that when the lockdown was implemented consecutively, the transmission rate was decreased, whereas, the recovery rate was found to be increased during the lockdown process. Reproduction number value was declining for the epidemic spread of Covid-19 data in India. Also after the strict governmental restrictions or strategies, the pandemic scene was found to be reduced drastically. This indicates that the awareness among people was increased much more than before during the lockdown period. Since there is good agreement between the stochastic simulation results using the estimated parameter values with the real Covid-19 data, the ABS SSA method is quite an important method for epidemiological parameter estimation. Further, these parameters can be used as a pandemic indicator as well as correlated to designing various pandemic control strategies.\\

{\noindent}Noise in the stochastic SIR model plays an important role in regulating pandemic/epidemic dynamics. The changes in dynamics can be captured by calculating Fano factor of the corresponding dynamics. We found that at the beginning of the pandemic/epidemic the dynamics suppresses noise effect and noise could not able to affect much (sub-Piossonian process). Once the dynamics crosses a transient state at which the dynamics becomes statistically uncorrelated (Poissonian state), the role of noise could able to drive the system to active pandemic/epidemic state (super-Poissonian or noise enhanced process). This noise or fluctuations in the epidemiological dynamics could be due to many factors, for example, diffusion of infected (symptomatic/asymptomatic or both) people in the demographic region (extrinsic noise) and emergence of people who do not follow epidemiological control strategies (WHO guidelines etc) enhancing disease spreading (intrinsic noise). Hence, this noise parameter could be one important factor to be included in devising epidemic control strategies. Further, the epidemiological parameter $\gamma=\frac{\beta}{R_0}$ is also found to be quite sensitive to the epidemiological dynamics and needs to be properly correlated with the control strategies to be implemented. \\

{\noindent}The analysis of the epidemiological parameters as a function of lockdown parameter $l$ showed that these parameters exhibit power law behavior with $l$. Since this power law behavior can be interpreted as fractal nature of the system with $l$, people during the lockdown period are self-organized and as a consequence, there is a signature of controlling the pandemic systematically. Hence, these parameters are quite sensitive to the pandemic dynamics and states and can be correlated in designing control strategies.

\vspace{0.5cm}
\noindent {\bf Acknowledgements} \\
JB is financially supported by the Indian Council of Medical Research (ICMR), New Delhi, India. RKBS is financially supported by DBT-COE and JNU.\\ \\
\noindent {\bf Author Contributions:}\\
{\noindent}RKBS and JB conceptualized the work. JB, and RKBS did the analytical work. JB  did the simulation work and also prepared the figures. All authors wrote, read, analyzed the results and approved the final manuscript.\\

\noindent {\bf Additional Information} \\
\textbf{Competing interests:} \\ The authors declare no competing interests.

\end{document}